# Quantitative characterization of surface topography using spectral analysis

Tevis Jacobs[1], Till Junge[2], Lars Pastewka[2,3]

[1] Department of Mechanical Engineering and Materials Science, University of Pittsburgh, 3700 O'Hara Street, Pittsburgh, PA 15261, USA

[2] Institute for Applied Materials, Karlsruhe Institute of Technology, Engelbert-Arnold-Straße 4, 76131 Karlsruhe, Germany

[3] MicroTribology Center µTC, Fraunhofer IWM, Wöhlerstraße 11, 79108 Freiburg, Germany

**Abstract**

Roughness determines many functional properties of surfaces, such as adhesion, friction, and (thermal and electrical) contact conductance. Recent analytical models and simulations enable quantitative prediction of these properties from knowledge of the power spectral density (PSD) of the surface topography. The utility of the PSD is that it contains statistical information that is unbiased by the particular scan size and pixel resolution chosen by the researcher. In this article, we first review the mathematical definition of the PSD, including the one- and two-dimensional cases, and common variations of each. We then discuss strategies for reconstructing an accurate PSD of a surface using topography measurements at different size scales. Finally, we discuss detecting and mitigating artifacts at the smallest scales, and computing upper/lower bounds on functional properties obtained from models. We accompany our discussion with virtual measurements on computer-generated surfaces. This discussion summarizes how to analyze topography measurements to reconstruct a reliable PSD. Analytical models demonstrate the potential for tuning functional properties by rationally tailoring surface topography – however, this potential can only be achieved through the accurate, quantitative reconstruction of the power spectral density of real-world surfaces.

## 1 Introduction

### 1.1 The importance of surface roughness and of the power spectral density as a description of topography

The surface roughness of a part has tremendous influence on its functionality. This has been reviewed extensively, e.g. in Ref. [1] and elsewhere. Roughness affects not only surface properties – such as hydrophobicity[2], optical and plasmonic behavior[3], adhesion[4–6], friction and Casimir forces[7] – but also "bulk" properties, such as fracture toughness and fatigue resistance[8]. For this reason, surface treatments are commonly used to control surface finish, and dozens of reference standards exist to describe measurement techniques and desired characteristics (such as ASME B46.1, ISO 4287, ISO 25178 and SEMI MF1811).

The power spectral density (PSD) of a surface is a mathematical tool that decomposes a surface into contributions from different spatial frequencies (wavevectors). Mathematically, the PSD is the Fourier transform of the autocorrelation function of the signal, which contains just the power (and not the phase) across a range of wavevectors [9–11]. This allows identification of the spatial frequencies that can be found in the signal. Figure 1 illustrates the PSD schematically for a regular dual-sinusoidal surface, a self-affine fractal surface, and a non-self-affine frozen capillary wave surface. Overall, the primary utility of the PSD is that it contains statistical information on the surface topography – which is largely unbiased by the choice of a particular scan size and pixel resolution picked by the researcher.

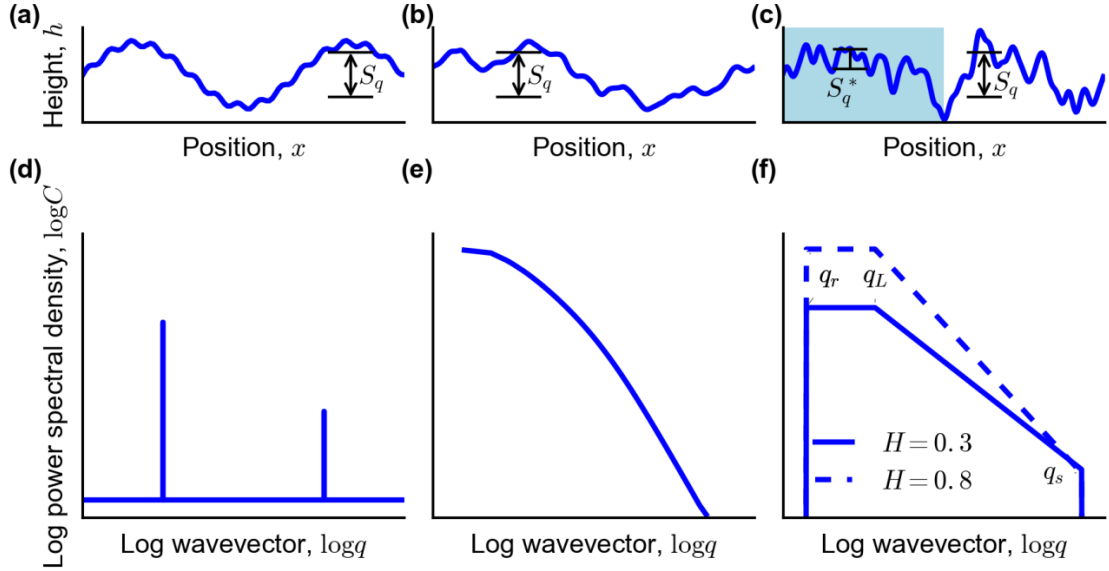

**Figure 1: Examples for regular and random surfaces.** Row (a)-(c) shows a line scan of a surface and (d)-(f) illustrates the corresponding power spectral density (PSD). Panel (a) shows a superposition of two sine waves of different wavelengths (wavevectors), panel (b) a frozen capillary wave and panel (c) a self-affine, randomly rough surface. (d) For the sine waves, the PSD has a single peak at the respective wavevector. The PSDs of randomly rough surfaces, as shown here for a non-self-affine frozen-capillary-wave surface (e) and two self-affine surfaces (f), have contributions from many wavevectors. The scalar measure of root mean square height, $S_q$, is identical for these three surfaces, although their topography is clearly different. The value $S_q^*$ shown for the randomly rough surface is measured over half the domain length indicated by the blue rectangle. This illustrates that $S_q$ depends on the domain over which the measurement takes place. Panel (f) shows the PSD for two surfaces with different Hurst exponents (solid and dashed line) but identical RMS slope $h'_{\text{rms}} = 0.1$. $q_r$, $q_L$ and $q_s$ indicate roll-off, large and small wavelength cutoffs.

Many contact mechanics models[12] have been constructed to compute functional properties of surfaces as a function of roughness. Specifically, quantities such as contact stiffness, adhesion, and the true contact area (which is distinct from the apparent contact area and which has relevance to friction[13–15] and electrical[16] and thermal[17] conductance across the contact) can be computed from knowledge of the PSD. Many of these models (such as Persson's scaling theory for contact[18–20]) take as an input the entire PSD of the surface. Other models (such as Refs. [5,6,21–26]) use mathematical manipulations such that the only input parameters are three scalar quantities: the root-mean-square (RMS) height $h_{\text{rms}}$, the RMS slope $h'_{\text{rms}}$, and the RMS curvature $h''_{\text{rms}}$ - all of which are calculated most reliably from the PSD itself, as discussed in Section 1.2. No matter which type of contact mechanics model is used, the central challenge for the practical application of these models is the measurement and calculation of accurate, reliable PSDs of real-world surfaces.

## 1.2 Using the power spectrum to compute accurate values for root-mean-square height, slope, and curvature of a surface

In an ideal case of perfect knowledge of surface topography as a continuous map of heights $h(x, y)$ at in-plane position $x, y$ with zero mean, the values of $h_{\text{rms}}$, $h'_{\text{rms}}$, and $h''_{\text{rms}}$ can be computed directly from the real-space topography. In this case, the root-mean-square height is given by $h_{\text{rms}}^2 = \langle h^2 \rangle$ where the angle brackets denote $\langle f \rangle = A_0^{-1} \int f(x,y) \mathrm{d}x \mathrm{d}y$, i.e. the average over the *x-y* plane of the enclosed function $f$. The RMS slope is computed as $h_{\text{rms}}'^2 = \langle |\nabla h|^2 \rangle$ and and the RMS curvature is computed as $h_{\text{rms}}''^2 = \frac{1}{2}\langle (\nabla^2 h)^2 \rangle$. Further, for many contact properties, just *one* of these three quantities will dominate the behavior. Specifically, the contact stiffness of an interface between two rough surfaces is primarily dependent on roughness at the large scale – and can be quantitatively related to the root-mean-square height $h_{\text{rms}}$ [22–25]. Analytical models and atomistic simulations show that other important quantities – such as the large-scale adhesive properties of a contacting junction[5,6,27], and the *true* contact area[21,26,28] – depend only on the smallest scales of roughness and therefore can be related to $h_{\text{rms}}'$, and $h_{\text{rms}}''$. Note that $h_{\text{rms}}$, $h_{\text{rms}}'$, and $h_{\text{rms}}''$ are single-valued and unique for a particular surface assuming perfect knowledge of that surface.

However, the central problem[29] of applying these analytical contact models to real-world surfaces is that it is not possible to determine a perfect description of surface topography as a continuous map of heights. Rather, we can only experimentally measure heights at discrete points, with finite resolution and over some finite measurement region. It is therefore not possible to directly determine $h_{\text{rms}}$, $h_{\text{rms}}'$, and $h_{\text{rms}}''$ from real-space measurements.

From any real-space measurement, we can of course compute approximate values for RMS height (typically denoted $R_q$ if obtained from a line scan and $S_q$ if obtained from an area scan), RMS slope $S_{\Delta q}$, and RMS curvature $S_{\Delta^2 q}$. However, only in an ideal experimental measurement with infinite scan size and infinite resolution will these be representative of the surface (*i.e.* in the ideal case, $S_q \equiv h_{\text{rms}}$, $S_{\Delta q} \equiv h_{\text{rms}}'$, and $S_{\Delta^2 q} \equiv h_{\text{rms}}''$). In real experiments, the measured value of $S_q$ will differ from the true $h_{\text{rms}}$ in cases where the measured region is insufficiently large or the resolution is insufficiently fine to accurately sample the topography. In almost every case, the measured values of $S_{\Delta q}$ and $S_{\Delta^2 q}$ will differ from $h_{\text{rms}}'$, and $h_{\text{rms}}''$, because the true local slope and curvature are computed from a discrete set of height information which is subject to experimental noise. Noise at small scales affects gradients and higher-order derivatives more than absolute height measurement. Additionally, there are multiple (infinite) ways to discretize the gradient operator, $\nabla h$. Two common standard slope measures $S_{\Delta q}$ (ISO 25178-2) and $S_{\Delta q6}$ (ASME B46.1) use a first- and a sixth-order finite difference stencil for $\nabla h$, respectively. Both can introduce either smoothing or artificially sharp corners, which will lead to inaccuracies. Smaller measurements with higher resolution will result in values of $S_{\Delta q}$ and $S_{\Delta^2 q}$ that are closer to $h_{\text{rms}}'$, and $h_{\text{rms}}''$ for the surface, but processing and instrumental artifacts tend to be most severe when pushing the resolution limits of an instrument.

The resolution to the problem of an incomplete measurement of a surface is to reconstruct the power spectral density (PSD) of the surface, as completely and accurately as possible. Specifically, one can compute, report, and analyze the PSD across many length scales. In some cases, it is possible to reconstruct an entire, complete PSD using multiple measurements and multiple techniques. Even in cases where the complete surface PSD cannot be measured due to instrument limitations, the partial PSD is extremely useful. It enables the calculation of upper and lower bounds for $h_{\text{rms}}$, $h_{\text{rms}}'$ and $h_{\text{rms}}''$ under certain assumptions; it can be used to identify and/or mitigate instrumental artifacts; and it can be extrapolated (in some cases) to learn about a surface beyond the limits of the measurement technique.

### 1.3 Challenges for the accurate experimental determination of the PSD

The PSD has been fruitfully applied to surface topography measurements for decades[30–33]. Many useful reference texts (for example Ref. [9]) describe in detail the calculation of the PSD for a surface. There is also an international reference standard describing its calculation (SEMI MF1811). Nevertheless, the PSD is underutilized quantitatively (except for extracting the Hurst exponent, which is related to the fractal dimension of a surface (see Appendix A.3)), and is used inconsistently. We believe that there are three critical challenges that must be addressed to enable the use of the PSD as a quantitative tool for prediction of properties. We discuss them in the following sections and strive to establish strategies for computing, analyzing, and reporting quantitative PSDs.

> Challenge A: While the mathematical description of the PSD is well-defined, there are several variations in the way it can be computed from measured data – each of which will lead to meaningfully different numerical results.
>
> Challenge B: The *theoretical* PSD that is used in many mathematical descriptions of surfaces such as contact mechanical models, is complete, accurate, and describes an ensemble of surfaces or a surface of infinite spatial extent. By contrast, the *experimental* PSD that is computed from a single measurement is incomplete (bandwidth-limited), inaccurate (artifact-prone), and describes a single iteration of a surface, which is finite in extent.
>
> Challenge C: Finally, recent contact theories predict that surface properties such as contact area and adhesion depend strongly on the smallest-scale of roughness. The current state-of-the-art experimental characterization techniques contain instrumental and analysis artifacts at these scales, which must be detected and mitigated.

The following sections serve to explain these challenges and to present strategies on how to mitigate them. To demonstrate these, we have created virtual surfaces *whose structure we know down to the smallest-scale*. By comparing the input PSD that was used to create the surface to the output PSD that was measured from the surface, the results demonstrate the influence of bandwidth and other instrumental limitations on the measured PSD. The purpose of the present paper is to discuss and review the application of spectral analysis to real, experimental surfaces. This topic has particular relevance to the application of self-affine scaling models; however, the concepts discussed apply equally well to the characterization of all surfaces – whether self-affine over a certain range or not.

## 2   Challenge A: Variations in the mathematical definition of the power spectral density

Many experimental investigations report the PSD as a useful measure of roughness for surfaces. However, a review of recent AFM-based PSD measurements on advanced-technology surfaces demonstrates significant inconsistency in the ways that PSDs are defined, computed, and analyzed. Even the units of the reported PSD vary from $m^2$ (Refs. [34,35]) to $m^3$ (Ref. [36–38]) to $m^4$ (Refs. [39–41]) to "arbitrary units" (Refs. [42,43]). Furthermore, the PSD is frequently used qualitatively to distinguish surfaces[38,41] and, when used quantitatively, is typically only used to determine the fractal dimension[35,42]. There is variation among these investigations in the calculation and interpretation of this value as well.

Here, we define and distinguish the various functions that are all referred to in common usage as *the* "power spectral density". This section of the main text distinguishes conceptually between types of PSD; rigorous mathematical definitions are included in Appendix A. Each of the various types of PSDs is the Fourier transform of the height autocorrelation function. By virtue of the convolution theorem of Fourier analysis, this can be computed as the square of the Fourier transform of the surface height $h(x,y)$. However, this calculation can be applied in several different ways to 1D and 2D signals. To concretely distinguish the different varieties and demonstrate their mathematical differences, the calculation of the root mean square height $h_{\text{rms}}$ is shown for each one.

## 2.1 Conceptual understanding of the different curves that are called "PSD"

For a one-dimensional signal (e.g., a line-scan from a stylus profilometer, as shown in Fig. 2a), the PSD $C^{1D}$ is a symmetric, one-dimensional function defined in frequency space (Fig. 2b). The units of $C^{1D}$ are [m$^3$] and the units of $q_x$ are [m$^{-1}$], such that the area under the curve, which is equal to $h_{\text{rms}}^2$, has the correct units of [m$^2$]:

$$h_{\text{rms}}^2 = \frac{1}{2\pi} \int_{-\infty}^{\infty} C^{1D}(q_x) \mathrm{d}q_x \tag{1}$$

$C^{1D}$ can be computed for a single measurement of a signal, or averaged over multiple measurements. Because this curve is always symmetric about $q_x=0$, it is more commonly represented by only showing the $q>0$ region of the curve, which we designate $C^{1D+}$ (Fig. 2c). Note that this is the most commonly-reported form of the PSD. When computing the RMS roughness using $C^{1D+}$, the area under the curve must be multiplied by 2 in order to account for the region where $q<0$:

$$h_{\text{rms}}^2 = \frac{1}{\pi} \int_0^{\infty} C^{1D+}(q_x) \mathrm{d}q_x \tag{2}$$

Note that Eq. (1) and (2) assume that the mean of the signal is zero, because it is common to subtract the mean value from the data.

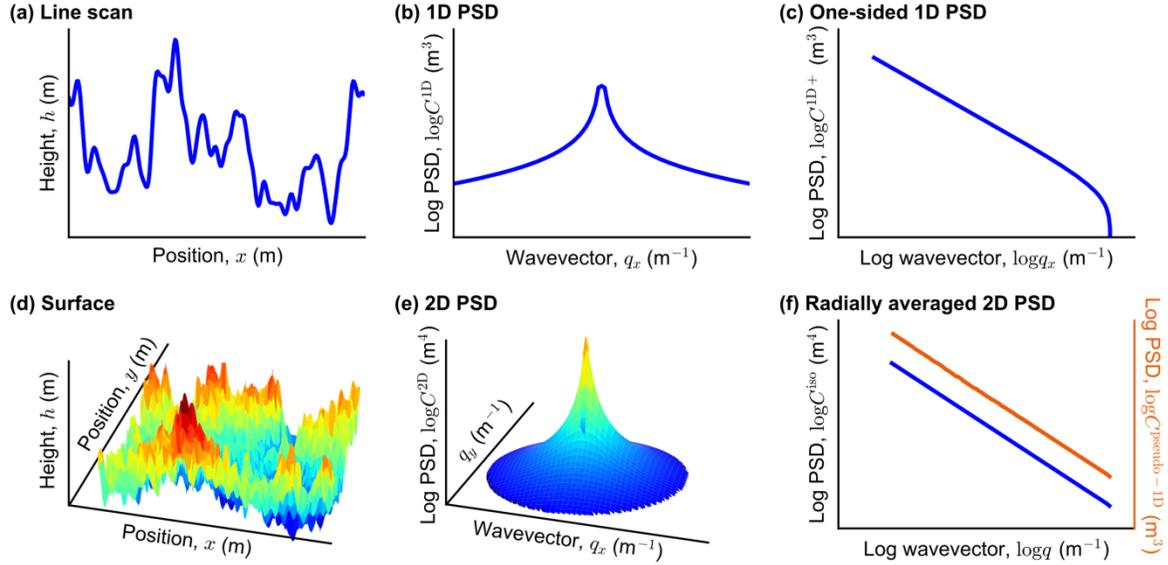

**Figure 2. Illustrating the various curves that are generally referred to as "power spectral density" (PSD).** A 1D line scan $h(x)$ (a) can be represented by a 1D PSD $C^{1D}(q_x)$ (b), but is more commonly indicated by a one-sided 1D PSD $C^{1D+}(q_x)$ (c) that omits wavevectors $q_x < 0$. A 2D topographic scan $h(x,y)$ (d) can be represented by a 2D PSD $C^{2D}(q_x, q_y)$ (e) that is a surface in reciprocal space $q_x$, $q_y$. For isotropic surfaces this function is radially symmetric and only (f) the radial average $C^{iso}(q)$ is typically shown. However, a pseudo-1D PSD is also commonly computed (according to Eq. (5) to enable easier comparison with $C^{1D+}$). Note that the units and absolute values are distinct between $C^{iso}$ and $C^{pseudo-1D}$.

By contrast, for a two-dimensional signal (*e.g.*, a topography map, as shown in Fig. 2d, the full two-dimensional PSD $C^{2D}$ is a surface in frequency space, shown in Fig. 2e. The units of $C^{2D}$ are [m$^4$] and the units of $q_x$ and $q_y$ are each [m$^{-1}$]. In this case, the volume under the PSD surface is equal to $h_{rms}^2$; this still has the correct units of [m$^2$]:

$$h_{rms}^2 = \frac{1}{4\pi^2} \int_{-\infty}^{\infty} \int_{-\infty}^{\infty} C^{2D}(q_x, q_y) dq_x dq_y \qquad (3)$$

While any vertical cross-section of this surface that passes through the origin *must* be symmetric with respect to inversion about $q_x=q_y=0$ because $h(x,y)$ is a real-valued function, the surface *need not* be radially symmetric with rotation about this point. In cases where it is radially symmetric – that is, where the real-space surface is isotropic – then the full PSD surface can be represented by taking a radial average about the origin. This yields the one-dimensional function $C^{iso}$ (Fig. 2f), which still has units of [m$^4$]. Converting $(q_x, q_y)$ to polar coordinates $(q, \theta)$ and integrating out $\theta$ yields a factor of $2\pi$ since $C^{iso}$ is invariant with $\theta$. The result is:

$$h_{rms}^2 = \frac{1}{2\pi} \int_0^{\infty} q C^{iso}(q) dq \qquad (4)$$

Many commercial software packages report a 1D PSD with units of [m$^3$] even for 2D surfaces because 1D PSDs are in more common usage; however, there are different ways to compute this 1D PSD from 2D data – and there can be differences in the resulting values. For instance, one method is to take the 1D PSD of each line of data, and then to take an average over all lines; we refer to this as $C^{1D+}$ for 2D data. Another method is to report a pseudo-1D PSD,

$$C^{\text{pseudo-1D}}(q) = \frac{q}{\pi} C^{\text{iso}}(q). \tag{5}$$

which follows $C^{\text{1D+}}$ at intermediate wavevectors for self-affine surfaces. However, $C^{\text{pseudo-1D}}$ cannot be treated like $C^{\text{1D+}}$ when computing scalar parameters. This is easily seen by inserting Eq. (5) into (4) to obtain

$$h_{\text{rms}}^2 = \frac{1}{2} \int_0^\infty C^{\text{pseudo-1D}}(q) \mathrm{d}q \tag{6}$$

which differs by a factor of $2/\pi$ from Eq. (2).

To complicate the picture, all of the previously described functions – $C^{\text{1D}}$, $C^{\text{1D+}}$, $C^{\text{2D}}$, $C^{\text{iso}}$, and $C^{\text{pseudo-1D}}$ – are imprecisely referred to as "the power spectral density" [34–42]. However, it is clear that they must be treated differently when quantitative values are computed. There are exact mathematical expressions relating the various forms to each other (see Appendix A.2 for more detail).

Most software packages are limited to reporting the one-dimensional representations $C^{\text{1D+}}$ or $C^{\text{pseudo-1D}}$ – as they are more amenable to conventional PSD analysis (most of which was developed for 1D time-series data). Indeed, the international reference standard SEMI MF1811 only discusses 1D profile measurements and some limited extension to 2D isotropic surfaces. These quantities are complete representations of the full $C^{\text{2D}}$ *only* in cases where the surface is isotropic. In the more general case, the mathematical integration over the full surface of $C^{\text{2D}}$ is required in order to compute accurate quantitative surface parameters. The mathematical use of $C^{\text{iso}}$ instead of $C^{\text{2D}}$ for a non-isotropic surface *will not* lead to errors in the value of $h_{\text{rms}}$ as computed in Eq. (4), but *will* lead to errors in $h'_{\text{rms}}$ and $h''_{\text{rms}}$ – which are discussed in the next section. Yet many software packages do not report the full $C^{\text{2D}}$ and may obscure the fact that the surface is anisotropic – leading to quantitative values computed from $C^{\text{iso}}$ that are unrepresentative of the surface.

### 2.2 Self-affine surfaces, and the resulting simplifications to the PSD

It has been shown that a wide range of natural and synthetic surfaces[44] – from coastlines[45] to mountain ranges[46,47] to fracture surfaces[48] to machined surfaces[49] – show characteristics of self-affine (also called *fractal*) scaling[30,32,50–52]. The underlying picture is that roughness consists of asperities (bumps), which are covered with smaller asperities, which in turn are covered with smaller asperities, etc. as was described by Archard in 1957.[12] The power spectral density (PSD) of a perfectly self-affine surface has a power-law dependence on the spatial frequency of roughness, and its exponent is related to the fractal dimension[32] of the surface (Fig. 1f). The calculations in Section 2.1 and Appendix A.1 and A.2 apply to all surfaces, regardless of self-affinity. However, in cases where self-affine roughness is observed, some of the calculations can be simplified.

Power spectra of real surfaces often show $C^{\text{iso}}(q) \propto q^{-2-2H}$ over many (but not all) scales [28,32]. This is the signature of a self-affine surface with Hurst exponent $H$ (see Appendix A.3 for more details). At small wavelength $\lambda$ (large wavevector $q$), the power-law will be cut off by the atomic spacing. The details of this cutoff are still debated. For the sake of the discussion in this paper, we assume that a sharp cutoff happens at a wavelength $\lambda_s$ (wavevector $q_s = 2\pi/\lambda_s$). At large wavelength $\lambda_L$ the power law typically crosses over to constant power, $C^{\text{iso}}(q) = \text{const}$ for $q < q_L = 2\pi/\lambda_L$. We assume that this "roll-off" can only extend up to a wavelength $\lambda_r$ (wavevector $q_r = 2\pi/\lambda_r$) where the power drops to zero. The idealized power-spectrum of a self-affine, randomly rough surface is therefore:

$$C^{\text{iso}}(q) = C_0 \begin{cases} 0 & \text{if } q < q_r \\ q_L^{-2-2H} = \text{const.} & \text{if } q_r \leq q < q_L \\ q^{-2-2H} & \text{if } q_L \leq q < q_s \\ 0 & \text{if } q \geq q_s \end{cases} \qquad (7)$$

where $q_r < q_L < q_s$ and $C_0$ is a constant. This power-spectrum is shown schematically in Fig. 1f. Shown there are two Hurst exponents (solid and dashed line) with identical RMS slope $h'_{\text{rms}} = 0.1$.

Self-affine scaling typically extends over many decades such that $q_s \gg q_L$. In this limit, we obtain simple analytical expressions for RMS slope and curvature (see also Appendix A.1):

$$(h'_{\text{rms}})^2 = \frac{1}{2\pi}\int_0^\infty q^3 C^{\text{iso}}(q)\,dq = \frac{1}{4\pi}\frac{C_0}{1-H}q_s^{2-2H} \qquad (8)$$

$$(h''_{\text{rms}})^2 = \frac{1}{8\pi}\int_0^\infty q^5 C^{\text{iso}}(q)\,dq = \frac{1}{16\pi}\frac{C_0}{2-H}q_s^{4-2H} \qquad (9)$$

It is important to note that because $q_s \gg q_L$, those expressions *do not depend* on $q_r$ and $q_L$. Slope and curvature are entirely determined by the structure at the smallest scales of the surface. Even if we did not have a sharp cutoff at $q_s$, the integral expressions that give $h'_{\text{rms}}$ and $h''_{\text{rms}}$ would be dominated by what happens at the smallest scales because of the power-law scaling of the PSD. This is graphically illustrated in Fig. 3.

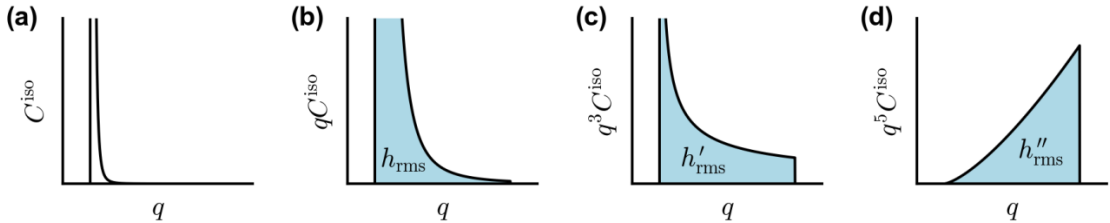

**Figure 3: Demonstration of the increasing contribution of the high-frequency content.** The calculation of $h_{\text{rms}}$ is relatively insensitive to errors in the value of the short-wavelength (high-wavevector) cut-off $q_s$ or the amplitude of $C^{\text{iso}}$ at that value. However, in the $h'_{\text{rms}}$ and $h''_{\text{rms}}$, the high-wavevector information becomes increasingly more important. Because surface properties such as contact area and macroscopic adhesion depend on the $h'_{\text{rms}}$ and $h''_{\text{rms}}$, the accuracy of calculations and predictions will depend on the accuracy of short-wavelength measurements. Note that unlike figures of other PSDs in this paper, this illustration shows all quantities on linear scales for both axes.

By contrast, the expression for RMS height depends on the power at the scale $q_L$ where the power-law region ends. With no roll-off $q_r \equiv q_L$ we get

$$h_{\text{rms}}^2 = \frac{C_0}{4\pi H}q_L^{-2H} \qquad (10)$$

while for a very large roll-off region $q_r \ll q_L$ we obtain

$$h_{\text{rms}}^2 = \frac{C_0}{4\pi}\left(1 + \frac{1}{H}\right)q_L^{-2H} \qquad (11)$$

Both expressions depend only on $C_0$ and $q_L$ and neither on $q_r$ nor on $q_s$.

We can analytically compute the PSD $C^{1D+}$ of a line profile that corresponds to the 2D, isotropic PSD given by Eq. (7). The resulting expression is (see Appendix A.3 for more detail):

$$\bar{C}^{1D}(q_x) \approx \frac{q_x}{\pi} C^{\text{iso}}(q_x)\sqrt{1-\left(\frac{q_x}{q_s}\right)^2}, \tag{12}$$

which forms the basis for the pseudo-1D PSD defined in Eq. (5). However, the 1D PSD will taper off approximately as $\sqrt{1-(q_x/q_s)^2}$ rather than drop to zero sharply. This behavior is illustrated in Fig. 2(c).

### 2.3 Strategies for the calculation of PSD and scalar roughness quantities

For the measurement and reporting of a PSD, we suggest the following guidelines. For quantitative numerical calculations such as in Refs. [21,53], $C^{\text{iso}}$ [m$^4$] is the correct form of the PSD to be used. However, the surface should also be checked for anisotropy, as most of the contact models assume isotropic roughness. For a 2D image of topography, this can be assessed by computing the surface PSD $C^{2D}$, in which the presence or absence of radial symmetry should be obvious, or by comparing $C^{1D+}$ against $C^{\text{pseudo-1D}}$, which should be identical for an isotropic surface.

If anisotropy is detected, it must be determined whether the surface is anisotropic or whether its origin is true anisotropy in the surface or artificial anisotropy introduced by the measurement technique, as is common for characterization techniques with a preferred direction. (For instance, in AFM the direction parallel to the scanning axis is sampled in less than one second, while topography in the perpendicular direction may take several minutes to measure: the latter is therefore much more prone to artifacts from drift.) This can be determined by repeating the measurement on the same surface in three different directions: horizontal, vertical and at a third, oblique direction. If the $C^{1D+}$ as calculated in these three directions is identical, then the surface is likely isotropic and, for self-affine surfaces, $C^{\text{iso}}$ can be approximated using $C^{\text{iso}}(q) = C^{1D+}(q)\pi/q$. If, on the other hand, $C^{1D+}$ is very different in the three directions, then the surface is likely to be truly anisotropic and isotropic contact models may not apply.

## 3 Challenge B: Reconstructing the theoretical PSD and scalar roughness parameters from incomplete data

In this section, we will distinguish among the PSD *of a measurement*, the PSD *of a surface*, and the PSD *of a class or an ensemble of surfaces*. One can always mathematically compute the PSD from any topography measurement carried out on a subsection of the full surface; however, the computed spectrum from any single measurement can only yield information for a limited range of wavelengths and will typically contain instrumental artifacts. Even ignoring instrumental artifacts (one of which is discussed in detail in Section 4), care is required to accurately reconstruct the full theoretical PSD to avoid artifacts due to: aperiodicity; artifacts of stitching; and small sample sizes. Finally, the calculation of upper and lower bounds of $h_{rms}$, $h'_{rms}$, and $h''_{rms}$ will be discussed.

To demonstrate each of these issues, we have analyzed computer-generated surfaces that were "measured" at various locations and resolutions. These synthetic measurements are, of course, free of measurement artifacts and therefore are useful for demonstrating the errors that can arise due to analysis procedure, rather than experimental procedure. These synthetic, computer-generated surfaces were created under the assumptions of the random process model of surface topography[30,50,51]. We start with a well-defined self-affine PSD of the form of Eq. (7), which we refer to as $C^{\text{iso}}_{\text{input}}(q)$, which is the *input PSD*. We then construct a surface as a superposition of waves of the form $h_{\vec{q}}(\vec{r}) = \chi_{\vec{q}} \exp(i\vec{q} \cdot \vec{r} + i\phi_{\vec{q}})$. The amplitude $\chi_{\vec{q}}$ is chosen randomly from a Gaussian distribution with a standard deviation given by the square-root of the PSD, $\left[C^{\text{iso}}_{\text{input}}(q)\right]^{1/2}$, for each $\vec{q}$. Each realization of a surface that corresponds to this input PSD will be subject to random (statistical) fluctuations. These fluctuations come from the random value of $\chi_{\vec{q}}$. We additionally chose a random phase shift $\phi_{\vec{q}}$ within the interval $[0, 2\pi)$ from a uniform distribution; however, this phase-shift does not affect the measured PSD. This way of creating synthetic PSDs is often referred to as the Fourier-filtering algorithm. All synthetic surfaces discussed here have a Hurst exponent of *H*=0.8.

### 3.1 Power spectral density of a single measurement – aperiodicity and the need for windowing

Experimental surfaces and their measurements will never be fully periodic. This introduces problems with a straightforward application of the Fourier transform described in Section 2 that treats the signal as periodic; thus, *windowing*[54] of the data is required [55]. The nonperiodic data is multiplied with a periodic windowing function that goes smoothly to zero at the edges of the topography image. In Fourier space, this window acts as a low-pass filter, removing the high-frequency components that are introduced by edges of the topography. While windowing is common practice in the signal-processing community, many conventional windows have a maximum that is set to unity. When using self-affine scaling laws to compute scalar roughness parameters, the sum rules that are used to compute $h_{\text{rms}}$, $h'_{\text{rms}}$ and $h''_{\text{rms}}$ must be conserved. Therefore, the area beneath the squared window *w* needs to be normalized to return the length *L* or area $L_x L_y$ of the window's support:

$$\int_0^L w^2(x)\mathrm{d}x = L \quad \text{or} \quad \int_0^{L_x}\int_0^{L_y} w^2(x,y)\mathrm{d}x\mathrm{d}y = L_x L_y. \tag{13}$$

Here we demonstrate the effect of aperiodicity, and artifacts that can arise from improper windowing using a realization of a well-defined random and statistically isotropic, periodic surface of $2048 \times 2048$ pixels that is shown in Fig. 4a. We assume the full surface is a 2μm × 2μm scan of a surface topography. The example surface used in this section has no power below a wavelength of 30 nm and has a Hurst exponent of 0.8.

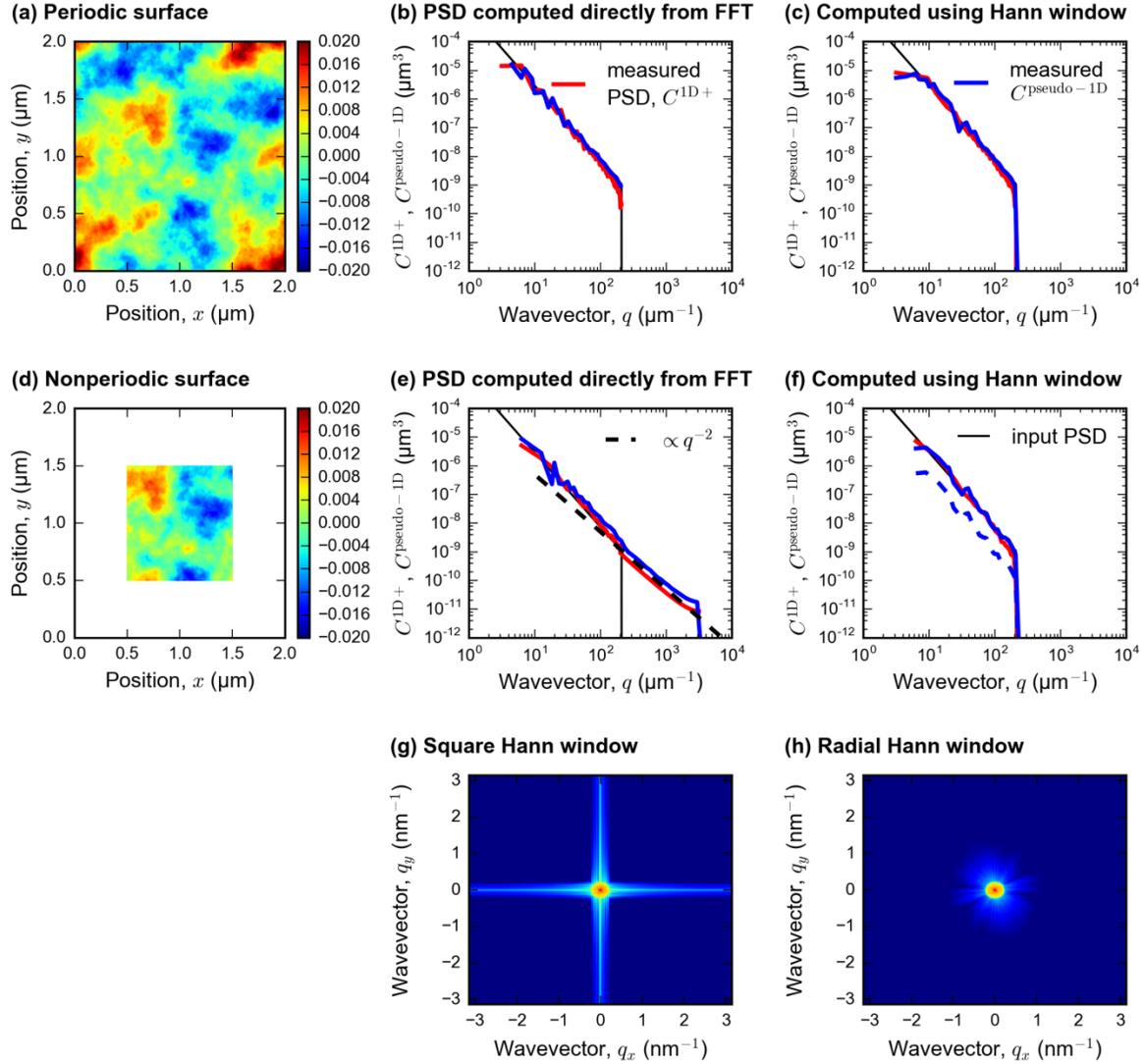

**Figure 4: Non-periodicity of measured data.** A synthetic periodic surface (a) containing 2048 × 2048 pixels was created with self-affine scaling properties and Gaussian statistics. The power-spectra are obtained using a straight FFT (b) and an FFT after applying a Hann window to the full topography (b). The black solid lines in show the input PSD that was used to generate these surfaces, and match well with measured values (red, blue lines). By contrast, the 1μm × 1μm central section of the topography (d) has been cropped from the surface shown in (a), and the resulting smaller surface is aperiodic. PSDs of the cropped region are then computed using a straight (e) and windowed (f) FFT. Without windowing, the PSD picks up an artificial contribution $\propto q^{-2}$ (dashed line in panel (e)) at large wavevectors (small wavelengths). The dashed line in panel (f) shows the PSD if a conventional normalization of the window is used, rather than normalizing to conserve the sum rules (Section 3.1). Further, the use of the square of a simple 1D Hann window (g) can introduce artificial asymmetry into the PSD; while the radial Hann window (h) faithfully reproduces the surface.

In the present paper, we exclusively use the Hann window[54], but results obtained with other windowing functions are virtually indistinguishable. The one-dimensional Hann window is given by

$$w^{1D}_{Hann}(x) = \left(\frac{2}{3}\right)^{1/2} (1 - \cos 2\pi x/L_x), \tag{14}$$

where $L_x$ is the length of the signal in the relevant direction (here designated $x$). Note that Eq. (14) fulfills the normalization conditions discussed above. In computing the 1D PSD, we apply this one dimensional window to each line scan independently, $h_{\text{windowed}}(x) = w_{\text{Hann}}^{1D}(x)h(x)$. Computing the 2D PSD requires a windowing function that is a function of $x$ and $y$ position within the plane. A common construction is to use the product of two one-dimensional windows, $w_{\text{Hann}}^{2D}(x,y) = w_{\text{Hann}}^{1D}(x)w_{\text{Hann}}^{1D}(y)$. However, the resulting window is not rotationally symmetric. Even if the surface is isotropic, the PSD of the windowed surface will become anisotropic, as shown in Fig. 4g. For these reasons, we use the radially symmetric Hann window,

$$w_{\text{Hann}}^{2D}(x,y) = \left(\frac{3\pi}{8} - \frac{2}{\pi}\right)^{-1/2} \left\{1 + \cos\left[\frac{2\pi\sqrt{X^2+Y^2}}{\min(L_x,L_y)}\right]\right\} \text{ for } X^2 + Y^2 < \left[\min(L_x,L_y)/2\right]^2, \quad (15)$$

where $X = x - L_x/2$ and $Y = y - L_y/2$ and the function is equal to zero everywhere the inequality is not satisfied.

Figure 4b shows the 1D PSD $C_{q_x}^{1D+}$ of the periodic surface shown in Fig. 4a, alongside $C_q^{\text{pseudo}-1D}$. These results show that the 1D and 2D PSDs for self-affine surfaces are equivalent, except for the region near the short-wavelength cutoff where the 1D PSD tapers off smoothly and the 2D PSD is cut off sharply [cf. Eq. (12)]. Both fall right on top of the input PSD shown by the black solid line. The PSD obtained with a Hann window, Fig. 4(c), follows the same power-law as the straight computation of the PSD and also recovers the same power in absolute terms. There are minor variations in the fluctuation of the PSD.

We now emulate a measurement of this surface by cutting out the central 1μm × 1μm section which is no longer periodic. The resulting surface is shown in Fig. 4d and the non-windowed and windowed PSDs are shown in Figs. 4e and 4f. It is clear that windowing is crucial to reconstructing the true PSD for non-periodic data. Critically, the non-windowed PSD makes the self-affine scaling appear to extend to the resolution limit of the measurement. In terms of windowing, the non-periodicity acts like a square window whose Fourier transform has a $1/q$ asymptotic behavior. The tail of the PSD in Fig. 4e therefore scales as $1/q^2$, as shown by the dashed line in that plot. Note that this looks like self-affine scaling with a Hurst exponent of $H=0.5$. For a real measurement of a real-world surface (where the true PSD is unknown), it could be erroneously concluded that self-affinity extends much further than it actually does. This artifact is particularly problematic when estimating the contact area using analytical models, as it would cause significant inaccuracy in the calculation of $h'_{\text{rms}}$ (and $h''_{\text{rms}}$), which in turn causes error in many predicted surface properties calculated from the PSD. Figure 4f shows that using the appropriate window resolves this issue.

Finally, Figs. 4g and 4h illustrate the effect of using a square and a radial window, respectively, on the 2D PSD. The window symmetry can clearly be seen in both images. The square window introduces apparent asymmetry into the 2D PSD: a vertical and a horizontal line of increased power. The use of a radial window does not bias the 2D PSD in any direction.

### 3.2 Power spectral density of a set of measurements

Mathematically and in simulations, the PSD of a surface can be perfectly understood over the whole frequency range. By contrast, the range of spectral information for a single measurement is limited[11] the maximum wavelength is determined by the size $L_x \times L_y$ of the domain over which the measurement was taken; the minimum wavelength is determined by the pixel size. The limit of *accurate* spectral information may be further reduced by instrumental artifacts[56], but this section is not considering instrumental artifacts. By combining multiple measurements, an individual technique is capable of providing spectral information over the range of wavelengths from the minimum instrument resolution to the maximum analysis size.

The most common measurement techniques employ scanning probes (i.e., stylus profilometry and atomic force microscopy) or light and/or x-rays (i.e, optical profilometry and angle-resolved scattering) to probe topography [57]. These techniques are summarized in Table 1. Stylus profilometry [58] has been used in some form for over 100 years and drags a sharp needle (typically 2-10 μm radius, but sometimes less than 1 μm) across a surface and records the deflection as a measure of surface topography. While the procedure is extremely robust and versatile, it is also relatively time-intensive, can cause sample damage, and can be limited by the relatively large tip. Atomic force microscopy (or, more generally, scanning probe microscopy) [59] is a class of techniques where a nanoscale tip is raster scanned over a surface. This tip can be in contact (contact mode), in intermittent contact (tapping mode), or out-of-contact (non-contact AFM or scanning tunneling microscopy). In all cases, the tip is used to sense the vertical position of the surface over an array of pixels in a grid. This technique can measure the smallest-scale features of any of the conventional surface characterization approaches and can achieve atomic resolution on ultra-flat samples. For rough samples, the lateral resolution is severely degraded by instrument noise and tip artifacts (see Section 3).

Optical profilometry [60] refers to a family of techniques that use either phase shifts of monochromatic light or optical coherence of white light to determine the vertical position of each pixel in an analysis region. Using analysis algorithms and fitting routines (such as those discussed in Refs. [61] and [62]), sub-nanometer resolution can be achieved in the vertical direction while the lateral resolution is diffraction limited to typical ranges of 500 nm–1 μm. The analysis size depends on the chosen microscope objective; while it is typically limited to approximately 5 mm for a single image, an encoded stage plus digital image registration can be readily used to stitch together multiple images to achieve 100 mm regions for analysis. Light scattering [9] and x-ray scattering [63] are a class of techniques that compare incident and scattered beams to measure changes in intensity as a function of wavelength. The details of this topic are beyond the scope of this paper, but unlike the other techniques discussed, scattering techniques do not measure the real-space topography of the sample, but rather using theories of wave-surface interactions [64] to relate the measured spectrum of the scattered beam to the spectrum of the surface. The scattered beam can be affected differently from different types of surfaces [65] and can be strongly affected by surface properties other than topography [63] and therefore various assumptions about the surface are often required. X-ray scattering can, in theory, sample roughness down to the wavelength of the x-ray (less than 1 Å), but in practice instrumental and measurement difficulties limit this to several nanometers.

**Table 1: Bandwidth limitations of various techniques for measuring surface topography. The ranges indicated cannot be achieved in a single measurement, but rather represent the ultimate limits of the techniques if multiple measurements are performed from the highest resolution through the highest scan size. Note that the frequency range for reliable data will be significantly reduced due to tip-size and noise artifacts, as discussed in Section 4.**

| Technique | Approx. maximum analysis size | Approx. lateral resolution limit | Frequency range | Advantages | Limitations |
|---|---|---|---|---|---|
| Stylus profilometry | 50 mm (200 mm with stitching) | 1 μm | $10^2$ ($10^1$) to $10^7$ m$^{-1}$ | Sub-nanometer height resolution. Unaffected by sample's optical properties | Contact method, can cause damage. Measurement is time-intensive. Tip shape can introduce artifacts |
| Optical profilometry | 5 mm (100 mm with stitching) | 1 μm | $10^3$ ($10^2$) to $10^7$ m$^{-1}$ | Rapid, non-contact. Sub-nanometer height resolution. | Diffraction-limited spatial information. Difficult to use on very rough surfaces. Artifacts from transparent thin films. |
| Light and x-ray scattering | 5 mm | 10 nm | $10^3$ to $10^9$ m$^{-1}$ | Rapid, non-contact. Insensitive to vibration. | Relies on models, assumptions to relate scattered beam to surface spectrum. Does not recover real-space topography |
| Scanning probe microscopy | 100 μm | 1 Å | $10^5$ to $10^{11}$ m$^{-1}$ | High-resolution | Small areas. Sensitive to vibration. Tip shape can introduce artifacts. |

To recover the complete PSD of a surface is typically necessary to combine the results from various techniques and multiple measurements per technique that are all performed on the same surface. The resultant PSDs can be stitched together over many orders of magnitude[11,47,66]. However, the process of stitching can introduce artifacts into the final PSD which cause a lack of overlap between scales[11]. To demonstrate this, we have created a very large (65,536 pixels per side) computer-generated surface that has a well-defined PSD. We use the central 50,000 pixels per side section of this surface as our realization of a 100-micron by 100-micron surface. The surface has then been "measured" (see Figs. 5a-c) with scan sizes down to 1 micron on a side. Each "measurement" had 500 by 500 pixels, therefore the pixel size scaled with scan size – as is common in measurement techniques such as AFM and optical profilometry. The measurement of the images was carried out by picking a random height value within each subpixel; though other ways to "sample" each pixel were also tested (i.e. picking subpixels at equal distances and finding the highest point in each pixel), but these details did not affect the results.

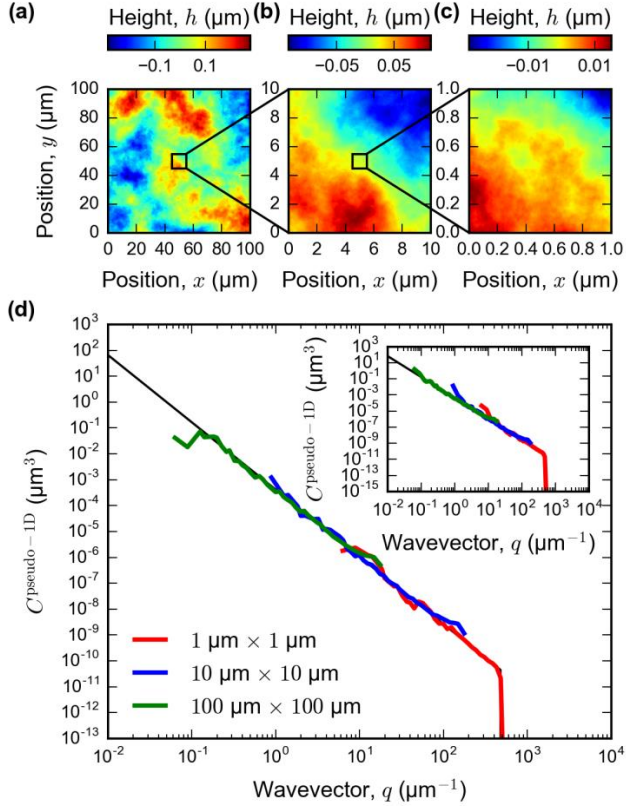

**Figure 5: Stitching PSDs and reproducibility of measurements.** A very large synthetic self-affine surface of 50,000 × 50,000 pixels has been digitally "measured" in blocks of 500 × 500 pixels with a variety of scan sizes ranging from the full surface with 100 microns width (a) to 10 microns (b) to 1 micron (c), with pixel sizes scaling accordingly. The individual PSDs are computed and plotted (d) after applying tilt equilibration and a Hann window independently to each emulated measurement. These are compared with the input PSD (black line) that was used to generate the surfaces. The inset shows the result without tilt correction, and demonstrates artifacts that cause a lack of overlap between the lowest frequencies of one measurement and the highest frequencies of the next.

Figure 5d shows the PSD stitched from three individual measurements at 100 μm, 10 μm and 1 μm image size taken in the middle of the surface. These results show first that sub-images must be tilt-compensated to eliminate errors at low $q$, as are evident in the inset image. The surfaces are untilted by subtracting an inclined plane $h(x, y) = ax + by + c$ to the topography such that the average slope of the surface is zero. While such a tilt compensation is commonplace in AFM measurements, there are other measurements where it is not typically done. An example is the technique recommended in Ref. [55] where one very large surface measurement is taken using optical reflectometry - then this large measurement is divided into sub-images for individual analysis. Even in this case, the sub-images must be tilt compensated so that each one is flat. When the sub-images are correctly windowed and correctly tilt-compensated, the overlay of PSDs is nearly perfect.

Finally, experimental measurements will almost always be limited to a range of wavevectors, such that the "true" surface PSD can never be fully determined. To account for this, any surfaces that show self-affine scaling over all or part of the spectrum can be fit with a power-law. This enables the use Eqs. (8)-(11) to compute $h_\text{rms}$, $h'_\text{rms}$, and $h''_\text{rms}$ for known bounds on long $q_L$ and short $q_s$ wavelength cutoffs. For example, a lower bound on $q_L$ can be obtained from the measurement at the lowest resolution (if the PSD levels off) and an upper bound to $q_L$ is $2\pi/L$ where $L$ is the maximum length of the geometry of the surface under investigation. Similarly, a lower bound on $q_s$ is $2\pi/a_0$ where $a_0$ is some characteristic interatomic spacing. The upper bound on $q_s$ is given by the resolution of the instrument used for the highest resolution topography measurement; a possible measure for scanning probe techniques is described in Section 4. These upper- and lower bounds for the cutoffs directly lead to upper and lower bounds on the scalar roughness parameters. Ideally, these bounds should be reported rather than a single value in all roughness investigations.

### 3.3 Strategies to reconstruct a multi-scale PSD

In light of the various factors discussed in this section, it is recommended that a tilt correction and a window (in this order) always be applied to every PSD measurement. We recommend that conclusions never be drawn about a surface from a single topography measurement. Surfaces may not be self-affine over all length scales and, even those that are, will have high and low-frequency cutoffs that affect the calculation of roughness parameters. Rather, it is recommended to compute a "master PSD" of the surface by combining PSDs from many topography measurements over as large a frequency range as possible. Not only should this master PSD include multiple surfaces and multiple locations per surface, but it should also include multiple different analysis techniques and a variety of sampling sizes for each technique. Finally, once this master PSD is computed, it should be used to compute upper and lower bounds for $h_{rms}$, $h'_{rms}$, and $h''_{rms}$. These will serve as a guide about whether more precise measurements are required for a given surface, and also will enable the calculation of uncertainty in surface properties predicted from contact models.

## 4 Challenge C: Accurately measuring topography at the smallest scales

Every measurement technique introduces artifacts in the measurement. Here we focus on artifacts introduced by scanning probe microscopy, because it provides the highest-resolution information of any conventional surface measurement technique and is commonly used to characterize surfaces at the nanoscale. AFM introduces many artifacts[67] into the measured data, including tip size effects, drift, acoustic and electronic noise, and image bow. In the context of prediction of surface properties, the most significant is the effect of tip size[68]. The purpose of this section is: to introduce the particular significance of this problem with respect to scalar roughness parameters; and to show the effect of tip-size on our synthetic surfaces and discuss practical upper limits on the frequency range that can be measured reliably.

### 4.1 Limitations of scanning probe microscopy for rough surfaces at the small scale

The finite size of the tip presents two related, but distinct problems: tip convolution; and feature deletion/creation. Tip convolution is the apparent blunting of sharp features because of the moving point-of-contact between the feature and the finite-size tip[69]. For example, the AFM measurement of an infinitely sharp spike will yield an inverse image of the tip itself. This effect has been extensively investigated[70–72] and algorithms have been developed to mathematically compensate, and reconstruct the unconvoluted surface topography[73–76]. However, feature deletion/creation occurs when the bluntness of the tip prevents the sampling of topography that is smaller than a certain spatial wavelength, and/or creates kinks in the data when there is a trench that is too narrow for the tip to reach the bottom[77]. While it is common in hard materials, it can also occur for soft samples when the surface fully conforms to the blunt tip, such that roughness features below a characteristic size cannot be sampled[78]. In these cases, there is no mathematical route to reconstruct the original surface from the measured data. The surface has not been accurately sampled and is indistinguishable from a range of other surfaces with the same problem.

At present, there is no simple solution to solving the high-frequency problem for atomic force microscopy. However, the investigator *can* determine the maximum frequency that can be reliably measured for a certain technique on a certain surface. This enables the determination of uncertainty on any quantities computed from the PSD.

The radius of the AFM probe determines the maximum spatial frequency that can be measured[68,79,80]. For a simple sine wave surface, the minimum wavelength $\lambda_c$ that can be sampled by a spherical-ended tip of radius $R_{tip}$ is given by [77,81,82]:

$$\text{minimum wavelength} = \lambda_c = 2\pi\sqrt{\chi R_{tip}} \tag{16}$$

where $\chi$ is the amplitude of the sine wave.

Church and Takacs [77] estimate more generally that the finite tip size leads to an asymptotic behavior of the power spectrum that follows,

$$C^{iso}(q) = C_{tip}q^{-5} \quad \text{or} \quad C^{1D}(q_x) = \frac{C_{tip}}{\pi}q_x^{-4}. \tag{17}$$

In the context of the PSD, a critical wavevector $q_c$ can be defined at which the contribution from tip curvature takes over and hides the true topography. This transition occurs at the wavevector where RMS curvature and tip curvature are approximately equivalent,

$$h''_{rms}(q_c) \approx c/R_{tip}, \tag{18}$$

where $c$ is a constant of order unity[77]. This criterion can be applied to a self-affine surface by combining Eq. (18) and Eq. (9) to yield

$$q_c = \left[\frac{16\pi c^2(2-H)}{C_0 R_{tip}^2}\right]^{\frac{1}{4-2H}} \tag{19}$$

This value of $q_c$ provides a *reliability cutoff* for AFM measurements. Data obtained at wavevectors larger than this value of $q_c$ are not reliable and no meaningful information about the PSD can be obtained. The prefactor of the tip-induced PSD is given by

$$C_{tip} = C_0 q_c^{3-2H}. \tag{20}$$

To demonstrate Eq. (20) we have scanned a synthetic surface of 0.5 μm × 0.5 μm size and lower wavelength cutoff $\lambda_s = 5$ nm with a spherical tip of radius 40 nm. For each in-plane position *x, y*, we lowered the tip onto the surface until the first point on the tip surface touched the rough specimen surface. Original and measured surfaces are shown in Fig. 6a and 6b, respectively. The resulting PSD is shown in Fig. 6c. It shows the proper power-law scaling down to $q_c$ where tip-shape effects take over. The dashed line shows the power-law given by Eq. (20) with $c = 1/2$ and a prefactor $C_0$ adjusted such that Eq. (18) is fulfilled numerically. It sits right on top of the simulated measurement, indicating that the analytical expression does indeed give the correct asymptotic behavior for a given tip radius. Data points in and near the blue region can therefore not be trusted in this measurement. If an investigator naively computed scalar roughness parameters from the measured PSD, the results would be in error by 9% for $S_{\Delta q}$ and by 27% for $S_{\Delta^2 q}$.

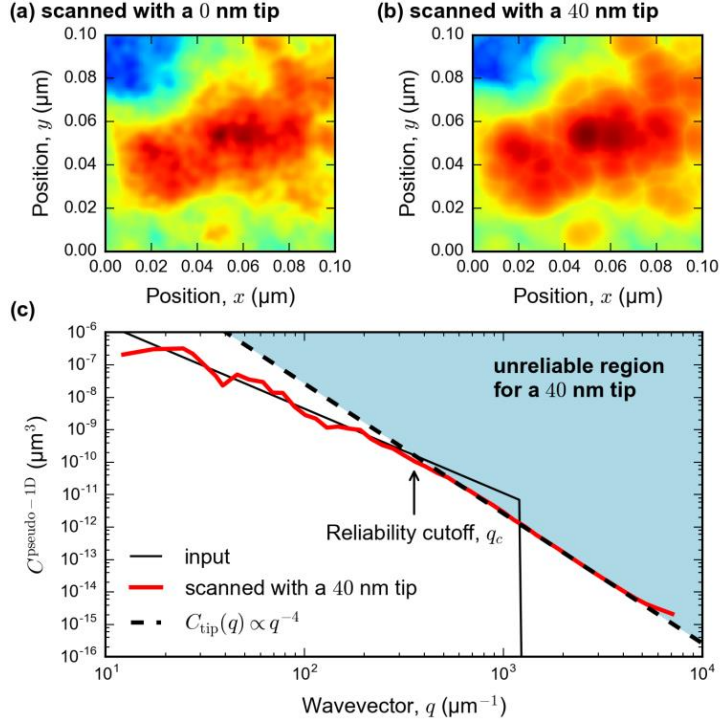

**Figure 6. Effect of tip radius.** A 0.1-µm square synthetic surface containing 200 pixels per side has been created from a subsection of a periodic self-affine surface 0.5 µm in size. When the surface is scanned with an infinitely sharp tip (a), the original surface is recovered. When the surface is scanned with an ideal spherical tip of radius 40 nm (b), some blunting is readily apparent. The blunted surface is used (after the application of a Hann window) to compute a PSD (d). The PSD of the scanned surface shows a $q^{-4}$ noise term at large wavevectors (small wavelengths). The asymptotic behavior is given Eq. (17) and shown by the dashed line. The black solid line shows the input PSD that was used to generate these surfaces. Note that the transition from real data to artifact-prone data is not readily visible in the computed PSD. Rather it is up to investigator to determine the reliability cutoff – using Eq. (19) - and to discard any PSD data for $q$ above this cutoff.

### 4.2 Effect of instrumental white noise

Each measurement instrument introduces noise into the process. The source of noise is manifold and ranges from thermally-induced oscillations in cantilever-based measurements [83] to thermal and shot noise in the measurement electronics [84]. We here test the influence of noise on the measurements of PSD by simply assuming uncorrelated white noise. This is an appropriate model for many sources of noise, such as thermal or shot noise.

White noise has a constant, wavelength-independent PSD. We introduce it into our synthetic surfaces by adding a random height, drawn from a Gaussian distribution with standard deviation $\sigma_{\text{noise}}$, to each topographic point. From Eq. (30) we immediately see that this leads to a constant contribution to the PSD with power

$$C_{\text{noise}}^{\text{iso}} = l_x l_y \sigma_{\text{noise}}^2 \qquad (21)$$

$l_x l_y$ is the pixel area; $l_x = N_x/L_x$ and $l_y = N_y/L_y$ where $N_x$ and $N_y$ is the number of grid points in $x$- and $y$-direction, respectively. For a one-dimensional PSD the expression becomes

$$C_{\text{noise}}^{1D} = l_x \sigma_{\text{noise}}^2. \tag{22}$$

It is interesting to note that the power depends not just on amplitude but also explicitly on measurement resolution. However, $\sigma_{\text{noise}}$ is a length which depends on the details of how the measured signal is converted into height information; this conversion may itself be affected by measurement resolution, care must therefore be taken in interpreting Eqs. (21) and (22) with respect to their scaling with resolution.

To demonstrate the effect of white noise, we calculate the PSD of a synthetic surface with additional noise consisting of $2000 \times 2000$ pixels. For each pixel $i,j$ we draw a random height $\Delta h_{ij}$ from a Gaussian distribution with standard deviation $\sigma_{\text{noise}}$ and add $\Delta h_{ij}$ to the topography map. The corresponding PSDs are shown in Fig. 7; $\sigma_{\text{noise}}$ was set to 20% of the root mean square height $h_{\text{rms}}$. The dashed lines in Fig. 7 show that white noise appears as a region of constant power for both 1D and 2D PSDs (panels (a) and (b), respectively) that is well-described by Eqs. (21) and (22). When the resolution is decreased to $200 \times 200$ pixels, the power of the noise increases as predicted. Note that for the low-resolution scan the noise intersects the power-law of the self-affine surface at a wavelength that is larger than the cutoff. The PSD then tapers off smoothly to constant power.

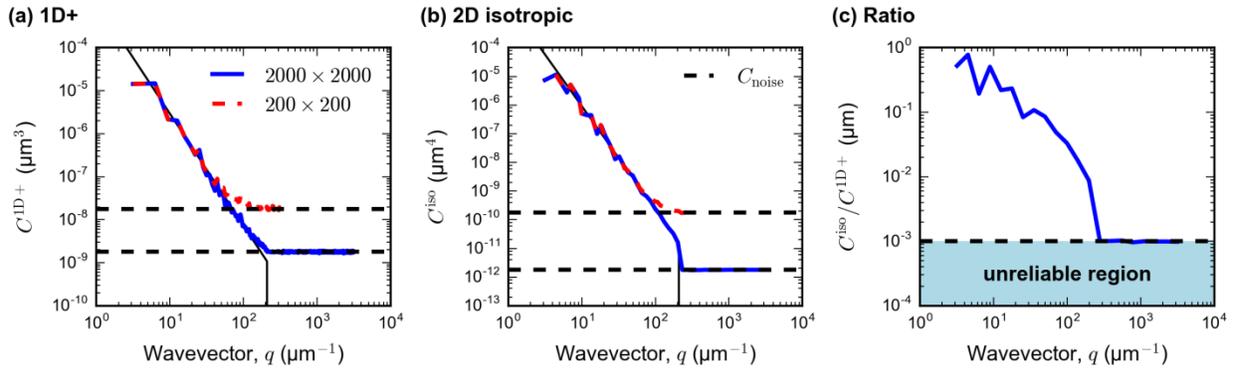

**Figure 7. Effect of instrumental white noise.** Blue lines show (a) 1D and (b) isotropic 2D PSDs of a periodic 2-μm square synthetic surface containing 2000 pixels per side. To mimic instrumental noise, we added random white-noise with standard deviation $\sigma_{\text{noise}} = 0.2\, h_{\text{rms}}$ to the synthetic topography image. Red lines show the PSD of the same surface but at a resolution of 200 pixels per side. The solid black line is the input PSD and the dashed lines are the noise amplitude given by Eqs. (21) and (22). Because surfaces are periodic, no window was applied before computing the PSD. Panel (c) shows the ratio of $C^{\text{iso}}/C^{1D+}$ for the surface with 2000 pixels per side. The horizontal dashed line on this plot at $C^{\text{iso}}/C^{1D+} = l_y$ indicates a reliability cutoff, where data below and close to this line are affected by instrumental noise and should be discarded.

The fact that white noise leads to a region of constant power in 1D and 2D PSDs can be used to detect it. The ratio of Eqs. (21) and (22) should be equal to the scan resolution in y-direction, $l_y$. Figure 7c shows this ratio as a function of wavevector $q$ for our synthetic surface. The dashed line shows $C^{\text{iso}}/C^{1D+} \equiv l_y$. The wavevector at which the ratio $C^{\text{iso}}/C^{1D+}$ intersects this line is the limit of reliability of the noisy data. Values at higher $q$ are affected by instrumental noise.

### 4.3 Strategies to detect and mitigate small-scale artifacts

Care must be taken when measuring and analyzing spectral content at the smallest scales. Perhaps the most concerning aspect of this reliability issue is that PSDs can be mathematically computed for arbitrarily small scan sizes and correspondingly large wavevectors. In many cases, the data will appear normal and there will be no inherent indication of where this "reliability cutoff" lies. Further, the effect on the PSD after this cutoff is not predictable. Depending on the profile and the tip shape, it may smoothen or roughen the profile, and may change the shape of the PSD[77]. Therefore, when performing AFM, it is recommended to determine the tip radius (by direct imaging[40,85] or by tip-reconstruction numerical algorithms[71,75]) and then to use Eqs. (19) and (20) to determine the maximum frequency that can be accurately measured using that combination of tip and surface. Then the noise must be analyzed, either qualitatively by looking for a level-off or quantitatively by computing $C^{\text{iso}}/C^{\text{1D+}}$ to determine the limit of the measurement noise. Spectral content beyond these limits should not be trusted, nor used for analysis. One can therefore specify only a lower bound for the high-frequency cut-off – and all properties that are computed from the PSD should reflect the uncertainty in this value. If higher-resolution information is required, then sharper tips and lower-noise equipment must be used, but even the sharpest tips have a limit of approximately 3-5 nm radius. For most surfaces that are not atomically smooth, this limits the small-wavelength limit to >10 nm. For this reason, new approaches must be devised in the future for transcending the limits of conventional techniques and rough surfaces down to the smallest scales.

## 5  Conclusions

Recent analytical models and numerical simulations make predictions for functional properties, such as macroscopic contact properties (*e.g.,* stiffness, contact area, and adhesion), on the basis of the power spectral density (PSD) of a surface. We have demonstrated three significant experimental challenges that hinder the application of these models to real-world surfaces – along with strategies to mitigate each one. First, there are several different well-accepted methods for the calculation of the PSD that result in different quantitative values. We review these and show how they relate to each other, and then discuss the considerations of using each. For analytical models of contact between randomly rough surfaces, the two-dimensional PSD is the correct one to use – rather than the "pseudo-one-dimensional PSD" that is computed by many software packages and recommended by international standards. Second, the analytical theories assume knowledge of the PSD of an infinite surface across a wide frequency range, while experimental measurements are plagued by edge effects and are necessarily limited to a narrow frequency band. We review the use of windowing and stitching to reliably combine multiple measurements, and discuss the calculation of upper and lower bounds on the true values of the scalar roughness parameters for a surface. Third, common techniques for surface topography measurement provide inaccurate data in the highest-frequency regime. We show the consequences of this issue for contact models and discuss the calculation of a "reliability cut-off" – beyond which measured PSD data should not be trusted, analyzed, nor reported.

We have demonstrated these various considerations using computer-generated surfaces. This analysis has shown that various specific types of artifacts can be found in experimentally-measured PSDs:

- Not properly accounting for the aperiodicity of the data by using the appropriate windows introduces a component $\propto q^{-2}$ into the 1D PSD ($\propto q^{-3}$ in the 2D PSD). This artifact appears identical to self-affine scaling with a Hurst exponent $H = 0.5$, but can be avoided by computing the PSD with a properly normalized, radially symmetric window.
- Not properly accounting for surface tilt introduces an overestimation of the PSD at low wavevectors, causing a lack of overlap between PSD measurements from scans of different sizes. This artifact can be avoided through tilt-compensation when computing the PSD.

- Tip shape in AFM measurements introduces a spurious component $\propto q^{-4}$ into the 1D PSD ($\propto q^{-5}$ in the 2D PSD). This artifact can be partially mitigated by the use of sharper tips, but cannot be avoided entirely. We have proposed a parameter-free expression that allows AFM users to identify the region of the PSD that is unreliable and should not be reported or used for calculation.
- Instrumental noise leads to constant power at the highest $q$. This artifact is easy to detect since it occurs in both 1D and 2D PSDs. We have proposed an approach for detecting the noise limit from the ratio of 2D and 1D PSDs.

Taken together, this article provides theoretical and practical guidance for the application of analytical roughness models to real-world surfaces.

## Acknowledgments
We thank Adam Hinkle for useful comments on the manuscript. This material is based upon work supported by the National Science Foundation under award number CMMI-1536800 and the Deutsche Forschungsgemeinschaft under grant PA 2023/2.

## Appendix A

### A.1    Mathematical definition of the power spectral density

Many excellent references describe the calculation of PSD, such as Ref. [28]. We will review the salient features here to establish the conventions used in this paper, and also to rigorously demonstrate the origins of some of the mathematical inconsistencies between the different types of PSDs which were mentioned in Section 2.

We start by assuming that the surface topography is given by a continuous function $h(x,y)$ with in-plane position $x, y$ and periodicity $L_x$ and $L_y$ in both Cartesian directions. We note that the choice of a function $h(x,y)$ excludes surfaces with overhangs that are sometimes observed experimentally. Due to the periodicity of $h(x,y)$, the Fourier transform $\tilde{h}_{q_x,q_y}$ is nonzero at discrete lattice points only. This type of transform is commonly referred to as the Fourier *series*, but we will use the term transform throughout the rest of the document. (We denote continuous functions using a typical function designation: $h(x,y)$ and we will represent discrete functions using subscripts $h_{x,y}$ here and elsewhere in this manuscript.) The forward and inverse transforms in the plane are given by the expressions

$$\tilde{h}_{q_x,q_y} = \int_A h(x,y) e^{-i(q_x x + q_y y)} \mathrm{d}x \mathrm{d}y \tag{23}$$

$$h(x,y) = \frac{1}{L_x L_y} \sum_{q_x,q_y} \tilde{h}_{q_x,q_y} e^{i(q_x x + q_y y)} \tag{24}$$

where the integral is over the full area $A = L_x L_y$ of the periodic topography profile and the sum runs between negative and positive infinity in steps of $2\pi/L_x$ and $2\pi/L_y$ in *x* and *y*, respectively.

Some authors use a pre-factor of $1/\sqrt{A}$ in front of both transforms or a factor of $1/A$ in front of the forward transform. Using one of these other conventions will change some of the pre-factors in the equations throughout this paper. It does also influence the *units* of the Fourier transform and hence the values obtained for the PSD. It is therefore important to agree on a canonical expression for the forward and inverse transforms.

Using Eq. (23) gives units of the Fourier transform $\tilde{h}$ as [m³]. In most real-world situations $h(x, y)$ is not described as a well-defined continuous function, but rather is measured only at discrete sites on a regular rectangular grid with "pixel" size $l_x \times l_y$. Here $l_x = N_x/L_x$ and $l_y = N_y/L_y$ where $N_x$ and $N_y$ is the number of grid points in *x*- and *y*-direction, respectively. In this case, the integral becomes a sum over all lattice sites and the sum in the inverse transform runs (in steps of $\Delta q_x = 2\pi/L_x$) from $q_x = -2\pi/l_x$ to $2\pi/l_x$ rather than from $q_x = -\infty$ to $\infty$. Eqs. (23) and (24) then become the forward and inverse discrete Fourier transforms (DFT),

$$\tilde{h}_{q_x,q_y} = l_x l_y \sum_{x,y} h_{x,y} e^{-i(q_x x + q_y y)} \tag{25}$$

$$h_{x,y} = \frac{1}{L_x L_y} \sum_{q_x,q_y} \tilde{h}_{q_x,q_y} e^{i(q_x x + q_y y)} \tag{26}$$

The most common numerical *algorithm* to compute the DFT is the fast Fourier transform (FFT)[86]. Because it is pervasively used, the terms FFT and DFT are often used synonymously. Note that in algorithmic implementations of Eqs. (25) and (26), $l_x = l_y = 1$ and $1/L_x L_y = 1/N_x N_y$ where $N_x \times N_y$ are the dimensions of the DFT grid.

We now *define* the two-dimensional PSD as

$$C^{2D}_{q_x,q_y} = A^{-1} \left| \tilde{h}_{q_x,q_y} \right|^2 \tag{27}$$

$C^{2D}_{q_x q_y}$ has units of [m⁴]. Equation (27) removes all phase information from $\tilde{h}_{q_x,q_y}$, but retains its amplitude. Because the reciprocal space product $\tilde{g}_{q_x,q_y} = \tilde{h}^*_{q_x,q_y} \cdot \tilde{h}_{q_x,q_y}$ (here * is the complex conjugate) becomes a convolution, $g(x, y) = \int h^*(x - x', y - y') h(x', y') dx' dy'$ in real space, Eq. (27) is the Fourier transform of the height autocorrelation function.

We show below that this normalization in combination with Eqs. (23) and (24) makes the PSD independent of sample size and allows comparison of PSDs measured over different areas. Eq. (27) is compatible with the definition of the 2D PSD in SEMI MF1811. Theoretical work by Persson uses a different normalization of the transform and the PSD, such that:

$$C^{\text{Persson}}_{q_x,q_y} = C^{2D}_{q_x,q_y} / 4\pi^2. \tag{28}$$

An important equality that derives from the properties of the convolution is Parseval's theorem, which relates the real-space power to the reciprocal-space power. Given the definition of the Fourier transform in Eqs. (23) and (24), Parseval's theorem has the form

$$\int_A |h(x,y)|^2 dx dy = A^{-1} \sum_{q_x q_y} |\tilde{h}_{q_x q_y}|^2 = \sum_{q_x q_y} C^{2D}_{q_x q_y}. \tag{29}$$

The PSD $C^{2D}_{q_x,q_y}$ contains certain scalar roughness parameters as simple sum rules. These are in particular the RMS roughness $h_{\text{rms}} = \sqrt{\langle |h|^2 \rangle}$, RMS slope $h'_{\text{rms}} = \sqrt{\langle |\nabla h|^2 \rangle}$ and RMS curvature $h''_{\text{rms}} = \frac{1}{2}\sqrt{\langle |\nabla^2 h|^2 \rangle}$. By virtue of Parseval's theorem, we find

$$h^2_{\text{rms}} = \frac{1}{A}\int_A h^2(x,y)\,dxdy = \frac{1}{A^2}\sum_{q_x,q_y} \left|\tilde{h}_{q_x,q_y}\right|^2 = \frac{1}{A}\sum_{q_x,q_y} C^{2D}_{q_x,q_y} \tag{30}$$

$$(h'_{\text{rms}})^2 = \frac{1}{A}\int_A |\nabla h(x,y)|^2\,dxdy = \frac{1}{A^2}\sum_{q_x,q_y}(q_x^2+q_y^2)\left|\tilde{h}_{q_x,q_y}\right|^2 = \frac{1}{A}\sum_{q_x,q_y} q^2 C^{2D}_{q_x,q_y} \tag{31}$$

$$(h''_{\text{rms}})^2 = \frac{1}{4A}\int_A |\nabla^2 h(x,y)|^2\,dxdy = \frac{1}{4A^2}\sum_{q_x,q_y}(q_x^2+q_y^2)^2\left|\tilde{h}_{q_x,q_y}\right|^2 = \frac{1}{4A}\sum_{q_x,q_y} q^4 C^{2D}_{q_x,q_y} \tag{32}$$

with $q = |\vec{q}| = \sqrt{q_x^2 + q_y^2}$.

It is useful to consider the equivalent expressions for a PSD for a surface of infinite extent where $\vec{q}$ becomes a continuous variable and to derive analytical expressions for $h^2_{\text{rms}}$ and $(h'_{\text{rms}})^2$ given an analytical expression $C^{2D}(q_x,q_y)$ for the PSD in this limit. For isotropic surfaces $C^{2D}_{q_x,q_y} = C^{\text{iso}}_q$ or $C^{2D}(q_x,q_y) = C^{\text{iso}}(q)$, and the PSD can be reported simply as a function of $q$. In many practical situations $|\tilde{h}_{q_x,q_y}|$ is a Gaussian random variable. (Indeed isotropy and normal distribution of $|\tilde{h}_{q_x,q_y}|$ is the basis of the random process model for surface topography[30,50,51] and hence also the basis for most modern theories of rough contact.) The isotropic PSD is obtained by averaging over all wavevectors that have $|\vec{q}| = q$, i.e., $C^{\text{iso}}(q) = (2\pi)^{-1}\int C^{2D}(q\cos\phi, q\sin\phi)\,d\phi$.

We can make the transition from discrete $C^{2D}_{q_x,q_y}$ to continuous $C^{2D}(q_x,q_y)$ by letting $A \to \infty$. Then for any function $f(q_x,q_y)$

$$\sum_{q_x,q_y} f(q_x,q_y) = \frac{1}{\Delta q_x \Delta q_y}\sum_{q_x,q_y} f(q_x,q_y)\Delta q_x \Delta q_y = \frac{A}{4\pi^2}\int f(q_x,q_y)\,dq_x dq_y \tag{33}$$

and Eqs. (30), (31) and (32) become

$$h^2_{\text{rms}} = \frac{1}{4\pi^2}\int C^{2D}(q_x,q_y)\,d^2q = \frac{1}{2\pi}\int q C^{\text{iso}}(q)\,dq \tag{34}$$

$$(h'_{\text{rms}})^2 = \frac{1}{4\pi^2}\int q^2 C^{2D}(q_x,q_y)\,d^2q = \frac{1}{2\pi}\int q^3 C^{\text{iso}}(q)\,dq \tag{35}$$

$$(h''_{\text{rms}})^2 = \frac{1}{16\pi^2}\int q^4 C^{2D}(q_x,q_y)\,d^2q = \frac{1}{8\pi}\int q^5 C^{\text{iso}}(q)\,dq \tag{36}$$

Note that these expressions are independent of area $A$ and hence well defined even in the limit $A \to \infty$.

### A.2 Relating the two-dimensional PSD to that of a line profile

One-dimensional power spectra are much more common than the previously discussed two-dimensional version due to their use in electrical engineering with time-varying electrical signals. For the same surface $h(x,y)$ the PSD of a line profile can be obtained from the one-dimensional Fourier series

$$\tilde{h}_{q_x}(y) = \int_{L_x} h(x,y) e^{-iq_x x} dx \tag{37}$$

$$h(x,y) = \frac{1}{L_x} \sum_{q_x} \tilde{h}_{q_x}(y) e^{iq_x x} \tag{38}$$

The 2D Fourier transform given in Eqs. (23) and (24) is simply the consecutive application of this formula in *x*- and *y*-direction. The 1D PSD is then given by

$$C_{q_x}^{1D}(y) = L_x^{-1} |\tilde{h}_{q_x}(y)|^2 \tag{39}$$

which has units of [m³] and depends explicitly on *y*, *i.e.*, the line of the scan. PSDs for line scans are typically reported as averages over multiple scans, denoted with a line over *C*,

$$\bar{C}_{q_x}^{1D} = L_y^{-1} \int C_{q_x}^{1D}(y) dy \tag{40}$$

Note that in the main text, the line over the *C* is dropped for brevity. By virtue of Parseval's theorem, we can express this averaged 1D PSD in terms of the 2D PSD,

$$\bar{C}_{q_x}^{1D} = L_y^{-1} \sum_{q_y} C_{q_x q_y}^{2D} \tag{41}$$

or for continuous $q_x, q_y$,

$$\bar{C}^{1D}(q_x) = \frac{1}{2\pi} \int_{-\infty}^{\infty} C^{2D}(q_x, q_y) dq_y \tag{42}$$

Note that a 2D power-spectrum can be reconstructed from 1D power spectra *if we assume that the surface is isotropic*, i.e. if $C^{2D}(q_x, q_y) = C^{iso}(q)$, with $q = \sqrt{q_x^2 + q_y^2}$ and $q dq = q_y dq_y$. Substituting *q* for $q_y$ in Eq. (42) gives

$$\bar{C}^{1D}(q_x) = \frac{1}{\pi} \int_{q_x}^{\infty} \frac{q C^{iso}(q)}{\sqrt{q^2 - q_x^2}} dq. \tag{43}$$

Inversion of this expression yields

$$C^{iso}(q) = -2 \int_q^{\infty} \frac{1}{\sqrt{q_x^2 - q^2}} \frac{d\bar{C}^{1D}(q_x)}{dq_x} dq_x \tag{44}$$

As discussed in the following section, for self-affine surfaces this is approximately equal to $\pi \bar{C}^{1D}(q_x)/q$.

### A.3 Power spectra for self-affine random surfaces

The two-dimensional PSD of a self-affine, randomly rough surface with Hurst exponent *H* follows the power-law expression $C^{iso}(q) \propto q^{-2-2H}$ (see also Eq. (7)). Self-affinity implies a certain scaling behavior that is encoded into this power-law. If all lengths are scaled by a factor of $\zeta$, then $q' = q/\zeta$ and

$$C^{iso}(q') d^2 q' = C^{iso}(q/\zeta) d^2 q/\zeta^2 = \zeta^{2H} C^{iso}(q) d^2 q, \tag{45}$$

i.e. heights need to be rescaled by $\zeta^H$ to give a surface with the same (statistical) roughness. This is the essence of self-affinity[32]. The Hurst[52] exponent $H$ of the power-law is related to the fractal dimension $D = 3 - H$. The values of $H$ are typically characteristic for the process that formed the surfaces. Typical values for $H$ on scales from atoms to mountains are between 0.7 and 0.9 (*e.g.*, Refs. [46,48]).

From Eqs. (34) to (36) we can easily derive the prefactor $C_0$ in $C^{\text{iso}}(q) = C_0 q^{-2-2H}$ [see also Eq. (7)] given either $h_{\text{rms}}$ and $q_L$, or $h'_{\text{rms}}$ and $q_s$. Under the assumption of scale separation $q_s \gg q_L$ we get:

$$C^{\text{iso}}(q) = 4\pi(1-H)\frac{(h'_{\text{rms}})^2}{q_s^4}\left(\frac{q}{q_s}\right)^{-2-2H} \tag{46}$$

$$C^{\text{iso}}(q) = 4\pi\alpha H \left(\frac{h_{\text{rms}}}{q_L}\right)^2 \left(\frac{q}{q_L}\right)^{-2-2H} \text{ with } \alpha = \begin{cases} 1 & \text{for } q_r = q_L \\ (1+H)^{-1} & \text{for } q_r \ll q_L \end{cases} \tag{47}$$

It is instructive to compute the PSD of a line scan of such an ideal isotropic and self-affine surface. We apply Eq. (43) to convert the two-dimensional PSD with $q_r = q_L$ into

$$\bar{C}^{\text{1D}}(q_x) = \frac{C_0}{\pi} q_x^{-1-2H} \; _2F_1\left(\frac{1}{2}; 1+H; \frac{3}{2}; 1-\left(\frac{q_s}{q_x}\right)^2\right) \sqrt{\left(\frac{q_s}{q_x}\right)^2 - 1} \tag{48}$$

where $_2F_1$ is the Gauss hypergeometric function. The expression takes a particularly simple form for $H = 1/2$ where $_2F_1(1/2, 3/2; 3/2; z) = 1/\sqrt{1-z}$. This yields

$$\bar{C}^{\text{1D}}(q_x) \approx \frac{C_0}{\pi} q_x^{-1-2H} \sqrt{1 - \left(\frac{q_x}{q_s}\right)^2} \tag{49}$$

and turns out to be an excellent approximation to Eq. (48) even for $H \neq 1/2$. The overall scaling is, as expected, $\propto q_x^{-1-2H}$ i.e. same as the two-dimensional PSD but with a power increased by one. The additional factor $\sqrt{1-(q_x/q_s)^2}$ tapers the function off to exactly zero at $q_x = q_s$. The deviation from $\propto q_x^{-1-2H}$ starts to become significant at about $q_x/q_s > 0.1$.

We can therefore express

$$\bar{C}^{\text{1D}}(q_x) \approx \frac{q_x}{\pi} C^{\text{iso}}(q_x)\sqrt{1-\left(\frac{q_x}{q_s}\right)^2}, \tag{50}$$

which forms the basis for the pseudo-1D PSD defined in Eq. (5).

## References

[1]   Bruzzone A A G, Costa H L, Lonardo P M and Lucca D A 2008 Advances in engineered surfaces for functional performance *CIRP Ann. - Manuf. Technol.* **57** 750–69

[2]   Dorrer C and Rühe J 2009 Some thoughts on superhydrophobic wetting *Soft Matter* **5** 51

[3]   Rodríguez-Fernández J, Funston A M, Pérez-Juste J, Álvarez-Puebla R A, Liz-Marzán L M and Mulvaney P 2009 The effect of surface roughness on the plasmonic response of individual sub-micron gold spheres *Phys. Chem. Chem. Phys.* **11** 5909

[4]     Delrio F W, de Boer M P, Knapp J A, Reedy Jr. D E, Clews P J and Dunn M L 2005 The role of van der Waals forces in adhesion of micromachined surfaces. *Nat. Mater.* **4** 629–34

[5]     Pastewka L and Robbins M O 2014 Contact between rough surfaces and a criterion for macroscopic adhesion *Proc. Natl. Acad. Sci. U. S. A.* **111** 3298–303

[6]     Müser M H 2015 A dimensionless measure for adhesion and effects of the range of adhesion in contacts of nominally flat surfaces *Tribol. Int.* 1–7

[7]     Van Zwol P J, Palasantzas G and De Hosson J T M 2007 Roughness corrections to the Casimir force: The importance of local surface slope *Appl. Phys. Lett.* **91** 1–3

[8]     Proudhon H, Fouvry S and Buffiere J 2005 A fretting crack initiation prediction taking into account the surface roughness and the crack nucleation process volume *Int. J. Fatigue* **27** 569–79

[9]     Stover J C 1995 *Optical Scattering: Measurement and Analysis* (SPIE Press)

[10]    Leach R K 2010 *Introduction to metrology for micro- and nanotechnology* (Elsevier)

[11]    Duparré A, Ferre-Borrull J, Gliech S, Notni G, Steinert J and Bennett J M 2002 Surface characterization techniques for determining the root-mean-square roughness and power spectral densities of optical components *Appl. Opt.* **41** 154

[12]    Archard J F 1957 Elastic Deformation and the Laws of Friction *Proc. R. Soc. A* **243** 190–205

[13]    Schwarz U D, Zwörner O, Köster P and Wiesendanger R 1997 Quantitative analysis of the frictional properties of solid materials at low loads. I. Carbon compounds *Phys. Rev. B* **56** 6987–96

[14]    Schwarz U D, Zwörner O, Köster P and Wiesendanger R 1997 Quantitative analysis of the frictional properties of solid materials at low loads. II. Mica and germanium sulfide *Phys. Rev. B* **56** 6997–7000

[15]    Carpick R W, Agraït N, Ogletree D F and Salmeron M 1996 Variation of the Interfacial Shear Strength and Adhesion of a Nanometer-Sized Contact *Langmuir* **12** 3334–40

[16]    Holm R 1967 *Electric Contacts* (Springer-Verlag Berlin Heidelberg)

[17]    Gotsmann B and Lantz M A 2013 Quantized thermal transport across contacts of rough surfaces *Nat. Mater.* **12** 59–65

[18]    Persson B N J 2001 Theory of rubber friction and contact mechanics *J. Chem. Phys.* **115** 3840–61

[19]    Persson B N J 2001 Elastoplastic contact between randomly rough surfaces *Phys. Rev. Lett.* **87** 116101

[20]    Peressadko A G, Hosoda N and Persson B N J 2005 Influence of surface roughness on adhesion between elastic bodies *Phys. Rev. Lett.* **95** 124301

[21]    Hyun S, Pei L, Molinari J-F and Robbins M O 2004 Finite-element analysis of contact between elastic self-affine surfaces *Phys. Rev. E* **70** 26117

[22]    Akarapu S, Sharp T and Robbins M O 2011 Stiffness of contacts between rough surfaces *Phys. Rev. Lett.* **106** 204301

[23]    Campañá C, Persson B N J and Müser M H 2011 Transverse and normal interfacial stiffness of

solids with randomly rough surfaces. *J. Phys. Condens. Matter* **23** 85001

[24]    Pohrt R and Popov V L 2012 Normal Contact Stiffness of Elastic Solids with Fractal Rough Surfaces *Phys. Rev. Lett.* **108** 104301

[25]    Pastewka L, Prodanov N, Lorenz B, Müser M H, Robbins M O and Persson B N J 2013 Finite-size scaling in the interfacial stiffness of rough elastic contacts *Phys. Rev. E* **87** 62809

[26]    Bush A W, Gibson R D and Thomas T R 1975 The elastic contact of a rough surface *Wear* **35** 87–111

[27]    Mulakaluri N and Persson B N J 2011 Adhesion between elastic solids with randomly rough surfaces: Comparison of analytical theory with molecular-dynamics simulations *EPL (Europhysics Lett.)* **96** 66003

[28]    Persson B N J, Albohr O, Tartaglino U, Volokitin A I and Tosatti E 2005 On the nature of surface roughness with application to contact mechanics, sealing, rubber friction and adhesion *J. Phys. Condens. Matter* **17** R1–62

[29]    Greenwood J A and Wu J J 2001 Surface Roughness and Contact: An Apology *Meccanica* **36** 617–30

[30]    Nayak P R 1971 Random process model of rough surfaces *J. Lubr. Technol.* **93** 398

[31]    Brown S R and Scholz C H 1985 Broad bandwidth study of the topography of natural rock surfaces *J. Geophys. Res.* **90** 12575

[32]    Mandelbrot B B 1982 *The Fractal Geometry of Nature* (W. H. Freeman)

[33]    Mandelbrot B B 1985 Self-Affine Fractals and Fractal Dimension *Phys. Scr.* **32** 257–60

[34]    Ulmeanu M, Serghei A, Mihailescu I ., Budau P and Enachescu M 2000 C–Ni amorphous multilayers studied by atomic force microscopy *Appl. Surf. Sci.* **165** 109–15

[35]    Bora C K, Flater E E, Street M D, Redmond J M, Starr M J, Carpick R W and Plesha M E 2005 Multiscale roughness and modeling of MEMS interfaces *Tribol. Lett.* **19** 37–48

[36]    Fang S J, Chen W, Yamanaka T and Helms C R 1995 Comparison of Si surface roughness measured by atomic force microscopy and ellipsometry *Appl. Phys. Lett.* **2837** 2837

[37]    Cui X, Hetke J F, Wiler J a, Anderson D J and Martin D C 2001 Electrochemical deposition and characterization of conducting polymer polypyrrole / PSS on multichannel neural probes *Sensors and actuators* **93** 8–18

[38]    Alsem D H, Xiang H, Ritchie R O and Komvopoulos K 2012 Sidewall adhesion and sliding contact behavior of polycrystalline silicon microdevices operated in high vacuum *J. Microelectromechanical Syst.* **21** 359–69

[39]    Karan S and Mallik B 2008 Power spectral density analysis and photoconducting behavior in copper(II) phthalocyanine nanostructured thin films *Phys. Chem. Chem. Phys.* **10** 6751–61

[40]    Senthilkumar M, Sahoo N K, Thakur S and Tokas R B 2005 Characterization of microroughness parameters in gadolinium oxide thin films: A study based on extended power spectral density analyses *Appl. Surf. Sci.* **252** 1608–19

[41]    Dash P, Mallick P, Rath H, Tripathi A, Prakash J, Avasthi D K, Mazumder S, Varma S, Satyam P V. and Mishra N C 2009 Surface roughness and power spectral density study of SHI irradiated ultra-thin gold films *Appl. Surf. Sci.* **256** 558–61

[42]    Sawant P D, Sabri Y M, Ippolito S J, Bansal V and Bhargava S K 2009 In-depth nano-scale analysis of complex interactions of Hg with gold nanostructures using AFM-based power spectrum density method. *Phys. Chem. Chem. Phys.* **11** 2374–8

[43]    Suzuki A, Yamazaki M, Kobiki Y and Suzuki H 1997 Surface Domains and Roughness of Polymer Gels Observed by Atomic Force Microscopy *Macromolecules* **30** 2350–4

[44]    Sayles R S and Thomas T R 1978 Surface topography as a nonstationary random process *Nature* **271** 431–4

[45]    Mandelbrot B B 1967 How Long Is the Coast of Britain? Statistical Self-Similarity and Fractional Dimension *Science* **156** 636–8

[46]    Gagnon J-S, Lovejoy S and Schertzer D 2006 Multifractal earth topography *Nonlinear Process. Geophys.* **13** 541–70

[47]    Candela T, Renard F, Klinger Y, Mair K, Schmittbuhl J and Brodsky E E 2012 Roughness of fault surfaces over nine decades of length scales *J. Geophys. Res. Solid Earth* **117** 1–30

[48]    Mandelbrot B B, Passoja D E and Paullay A J 1984 Fractal character of fracture surfaces of metals *Nature* **308** 721–2

[49]    Hasegawa M, Liu J, Okuda K and Nunobiki M 1996 Calculation of the fractal dimensions of machined surface profiles *Wear* **192** 40–5

[50]    Longuet-Higgins M S 1957 The statistical analysis of a random, moving surface *Phil. Trans. R. Soc. Lond. A* **249** 321–87

[51]    Greenwood J A 1984 A unified theory of surface roughness *Proc. R. Soc. A Math. Phys. Eng. Sci.* **393** 133–57

[52]    Meakin P 1998 *Fractals, scaling and growth far from equilibrium* (Cambridge University Press)

[53]    Campañá C and Müser M H 2007 Contact mechanics of real vs. randomly rough surfaces: A Green's function molecular dynamics study *EPL (Europhysics Lett.)* **77** 38005

[54]    Prabhu K M M *Window Functions and Their Applications in Signal Processing* (CRC Press)

[55]    Elson J M and Bennett J M 1995 Calculation of the power spectral density from surface profile data. *Appl. Opt.* **34** 201–8

[56]    González Martínez J, Nieto-Carvajal I, Abad J and Colchero J 2012 Nanoscale measurement of the power spectral density of surface roughness: how to solve a difficult experimental challenge *Nanoscale Res. Lett.* **7** 174

[57]    Leach R and Haitjema H 2010 Bandwidth characteristics and comparisons of surface texture measuring instruments *Meas. Sci. Technol.* **21** 32001

[58]    Whitehouse D J 2011 *Handbook of surface and nanometrology* (CRC Press)

[59]   Meyer E, Hug H J and Bennewitz R 2004 *Scanning Probe Microscopy: The Lab on a Tip* (Springer)

[60]   Groot P J De 2015 Interference Microscopy for Surface Structure Analysis *Handbook of Optical Metrology: Principle and Applications* (CRC Press) pp 791–828

[61]   de Groot P 2015 Principles of interference microscopy for the measurement of surface topography *Adv. Opt. Photonics* **7** 1

[62]   Tan J, Liu C, Liu J, Wang H, B T B V R and R, Al U G et, Li S G, Xu Z G, Reading I, Yoon S F F Z R and Z J, Köklü F H Q J I and V A N, Dumitriu D R A and M J H, Liu J, Wang Y, Liu C, Wilson T W H and T J, R W T and M B, Tan J L J and W Y, R L, Liu Y G, Liu L A F and Y X C, R S C J, R W T and S C J, M G and J M J 2016 Sinc2 fitting for height extraction in confocal scanning *Meas. Sci. Technol.* **27** 25006

[63]   Foster M D 1993 X-Ray Scattering Methods for the Study of Polymer Interfaces *Crit. Rev. Anal. Chem.* **24** 179–241

[64]   Ogilvy J A and Merklinger H M 1991 Theory of Wave Scattering from Random Rough Surfaces *J. Acoust. Soc. Am.* **90** 3382

[65]   Vorburger T V., Marx E and Lettieri T R 1993 Regimes of surface roughness measurable with light scattering *Appl. Opt.* **32** 3401

[66]   Marx E, Malik I J, Strausser Y E, Bristow T, Poduje N and Stover J C 2002 Power spectral densities: A multiple technique study of different Si wafer surfaces *J. Vac. Sci. Technol. B* **20** 31

[67]   Fang S J, Haplepete S, Chen W, Helms C R and Edwards H 1997 Analyzing atomic force microscopy images using spectral methods *J. Appl. Phys.* **82** 5891

[68]   Westra K L and Thomson D J 1995 Effect of tip shape on surface roughness measurements from atomic force microscopy images of thin films *J. Vac. Sci. Technol. B Microelectron. Nanom. Struct.* **13** 344

[69]   Villarrubia J S 1994 Morphological estimation of tip geometry for scanned probe microscopy *Surf. Sci.* **321** 287–300

[70]   Tranchida D, Piccarolo S and Deblieck R a C 2006 Some experimental issues of AFM tip blind estimation: the effect of noise and resolution *Meas. Sci. Technol.* **17** 2630–6

[71]   Flater E E, Zacharakis-Jutz G E, Dumba B G, White I A and Clifford C A 2014 Towards easy and reliable AFM tip shape determination using blind tip reconstruction *Ultramicroscopy* **146** 130–43

[72]   Dongmo L S, Villarrubia J S, Jones S N, Renegar T B, Postek M T and Song J F 2000 Experimental test of blind tip reconstruction for scanning probe microscopy *Ultramicroscopy* **85** 141–53

[73]   Williams P M, Shakesheff K M, Davies M C, Jackson D E, Roberts C J and Tendler S J B 1996 Blind reconstruction of scanning probe image data *J. Vac. Sci. Technol. B Microelectron. Nanom. Struct.* **14** 1557

[74]   Dahlen G, Osborn M, Okulan N, Foreman W, Chand A and Foucher J 2005 Tip characterization and surface reconstruction of complex structures with critical dimension atomic force microscopy *J. Vac. Sci. Technol. B Microelectron. Nanom. Struct.* **23** 2297

[75]    Villarrubia J S 1997 Algorithms for scanned probe microscope image simulation, surface reconstruction, and tip estimation *J. Res. Natl. Inst. Stand. Technol.* **102** 425

[76]    Xu M, Fujita D and Onishi K 2009 Reconstruction of atomic force microscopy image by using nanofabricated tip characterizer toward the actual sample surface topography *Rev. Sci. Instrum.* **80**

[77]    Church E L and Takacs P Z 1991 Effects of the nonvanishing tip size in mechanical profile measurements *Proc. SPIE 1332, Optical Testing and Metrology III: Recent Advances in Industrial Optical Inspection* vol 1332pp 504–14

[78]    Knoll A W 2013 Nanoscale Contact-Radius Determination by Spectral Analysis of Polymer Roughness Images *Langmuir* **29** 13958–66

[79]    O'Donnell K A 1993 Effects of finite stylus width in surface contact profilometry *Appl. Opt.* **32** 4922–8

[80]    Sedin D L and Rowlen K L 2001 Influence of tip size on AFM roughness measurements *Appl. Surf. Sci.* **182** 40–8

[81]    Church E L and Takacs P Z 1989 Instrumental effects in surface finish measurement *Proc. SPIE 1009, Surface Measurement and Characterization* vol 1009pp 46–55

[82]    Leach R K, Weckenmann A, Coupland J and Hartmann W 2014 Interpreting the probe-surface interaction of surface measuring instruments, or what is a surface? *Surf. Topogr. Metrol. Prop.* **2** 35001

[83]    Butt H-J, Jaschke M, Binnig G Q C F and G C, D S, G B, H M E and H, Martin Y W C C and W H K, S T, I S, Gould S A C, Drake B, Prater C B, Weisenhorn A L, Manne S, Kelderman G L, Butt H-J, Hansma H, Hansma P K M S and C H J, C S, W W, A K, W R J, J C, S B, Albrecht T R, Akamine S C T E and Q C F, Butt H-J, Siedle P, Seifert K, Fendler K, Seeger T, Bamberg E, Weisenhorn A L G K and E A, L S J E and W and J H J L and B 1995 Calculation of thermal noise in atomic force microscopy *Nanotechnology* **6** 1–7

[84]    Motchenbacher C D and Connelly J A 1993 *Low-Noise Electronic Design* (John Wiley & Sons, Inc.)

[85]    Jacobs T D B, Wabiszewski G E, Goodman A J and Carpick R W 2016 Characterizing nanoscale scanning probes using electron microscopy : A novel fixture and a practical guide *Rev. Sci. Instrum.* **87** 13703

[86]    Van Loan C 1992 *Computational Frameworks for the Fast Fourier Transform* (Society for Industrial and Applied Mathematics)